\documentclass{WileyMSP-template}
\usepackage{cite}
\usepackage{amsmath,amssymb,amsfonts}
\usepackage{algorithmic}
\usepackage{graphicx}
\usepackage{textcomp}
\usepackage{longtable}
\usepackage{tabularx,booktabs}
\usepackage{caption}
\usepackage{float}
\usepackage{dblfloatfix}
\usepackage{soul}
\usepackage{fancyhdr}
\begin{document}

\pagestyle{fancy}
\title{Teaching Undergraduate Students to Think Like Real-World Systems Engineers: A Technology-Based Hybrid Learning Approach}
\maketitle

\author{Rami Ghannam*}
\author{Cecilia Chan}

\begin{affiliations}
Dr. R. Ghannam. \\
James Watt School of Engineering, University of Glasgow.\\
Glasgow G12 8QQ, UK.\\
Email Address: rami.ghannam@glasgow.ac.uk
\\
\vspace{2mm}
Dr. C. Chan.\\
Centre for the Enhancement of Teaching and Learning. \\
The University of Hong Kong.\\
Email Address: cecilia.chan@cetl.hku.hk

\end{affiliations}


\begin{abstract}
A hybrid teaching approach that relies on combining the principles of Project Based Learning and Team Based Learning has been designed in an electronic engineering module during the past five years. Our motivation was to expose students to real-world authentic engineering problems and to direct students away from the classical “banking” teaching approach, with a view to developing their systems engineering skills via deeper and more collaborative learning. Adopting this hybrid teaching approach in an electronic engineering curriculum enabled students to integrate knowledge in areas that included control systems, image processing, embedded systems, sensors, as well as a combination of team working, decision making, trouble shooting and project planning skills. This article demonstrates how a hybrid learning approach that combines technology with active learning was used to design an innovative third year module called Team Design and Project Skills, which was concerned with students dividing themselves in teams to develop a smart electronic system. We reveal module design details that led to 320 students taking this systems engineering course. We also discuss the effectiveness of our teaching approach via analysis of student grades during the past five years, as well as data from surveys that were completed by 68 students. Results from our surveys showed that 64\% agreed that the module helped students broaden their perspective in electronic systems design. Moreover, 84\% recognised that this module was a valuable component in their degree programme. 
\end{abstract}

\section{Introduction}
\begin{figure}
  \includegraphics[width=\linewidth]{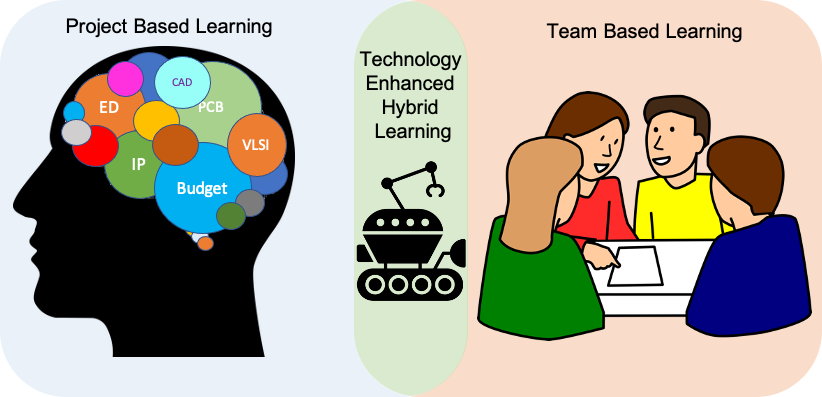}\\
  \caption{Concept of the proposed hybrid learning technique, which combines the elements of Project Based Learning (PBL) with Team Based Learning (TBL) via a technology enhanced learning activity.}
  \label{fig:HybridLearningConcept}
\end{figure}

The nature, complexity and type of problems that practicing engineers are required to solve are different from the classical text book problems that students are exposed to, since real-world problems are often ill-bounded and not well defined. Such problems require interdisciplinary knowledge and multiple solutions may exist, or no solutions at all \cite{Kitchner1983}. Moreover, engineers in the real-world typically solve problems in teams. Rarely do engineers work in solitude and collaboration is therefore necessary \cite{Jonassen2006}.  
\\
\vspace{5mm}
There has been an ongoing dialogue between academia, accreditation bodies and industry regarding the range of technical and interpersonal skills required from graduating engineers to meet the needs of the continuously shifting job market. According to the literature, engineers lack the necessary team working, communication, social and emotional graduate attributes needed in today's job market  \cite{InternationalCo-operationEuropeLtd.ICEL2001}. Therefore, engineering accreditation bodies now require universities to demonstrate that their students have enough opportunities to develop these skills during their engineering studies. To achieve the above objectives, the attributes of Team-Based Learning (TBL) with Problem Based Learning (PBL) were combined in a module, such that students ‘learn’ teamwork, collaboration, project management and communication as by products of the activity within the module (c.f. figure \ref{fig:HybridLearningConcept}). Instead of being taught these interpersonal skills, our approach relied on getting absorbed by a technology enabled challenge that makes them forget the time and effort spent on developing them. Therefore, students were given an opportunity to develop both interdisciplinary systems engineering knowledge as well as essential interpersonal skills in a single module.
\\
\vspace{5mm}
Furthermore, active learning techniques have demonstrated improved student understanding of engineering concepts \cite{Prince2004,Freeman2014}. Instead of the traditional teacher-centred instruction approach, our motivation was to encourage more modules to adopt a student-centred learning, such that students can re-use their skills across a range of different modules \cite{Miller2014}. In fact, the strategy adopted in this investigation involved optimizing student learning using a varied or systemic approach that fits well with real-world engineering practices \cite{baron2020adopting}. Such innovative teaching approaches are necessary, since it simply becomes infeasible to cover the rapidly evolving electronic engineering field in a typical 4-year undergraduate programme. Therefore, optimised student learning becomes essential.

\section{Research Context}
Project Based Learning (PBL) is a student-centred teaching strategy that has proven to improve student performance. It focuses on practical, real world problems that aim to increase student motivation. Typically, PBL involves splitting students into groups of 6 to 10 students to work on a project that is facilitated by a single instructor for six to ten weeks. Student groups are then shuffled to tackle another project \cite{Kokotsaki2016, Macias-Guarasa2006}. There are numerous examples of PBL adopted in electronic engineering disciplines for enhancing student learning and embedding soft skills \cite{Hosseinzadeh2012, Kellett2012, Kumar2013, Lee2010,Lee2016,Calvo2018,Hassan2015, Mantri2008}. A thorough review of project based learning (PBL) is available by \cite{Kokotsaki2016}. Evaluations of these interventions have focused almost exclusively on student interviews or responses to open-ended questions, which have found that students are in favour of courses that implement PBL in engineering programmes. 
\\
\vspace{5mm}
Most recently, Salankar \textit{et al.} investigated the impact of PBL on the performance of engineering students \cite{Salankar2021} using the OCEAN (Openness, Conscientiousness, Extraversion, Agreeableness and Neuroticism) personality model \cite{Durupinar2008}. Based on surveys that were completed by 77 students, Openness and Extraversion were personality traits that were significantly improved. Furthermore, the use of PBL in a robotics course was investigated by Calvo \textit{et al.} \cite{Calvo2018}. While the study provided a thorough description of the PBL activity, there was brief insight into how student satisfaction was measured, how the surveys were conducted nor the instruments used to conduct their investigation. There was also no mention of the percentage of students who took part in the surveys (response rate). Student satisfaction was measured using two surveys that elicited information about the teaching method, workload and course interest. However, a small sample size was used in each of the four years (8 to 38) and student choices were limited to only 4. Similarly, the authors in \cite{Martinez-Rodrigo2017} investigated student satisfaction for a PBL activity in a third year power electronics course. Their course consisted of 60 class hours and 90 hours of independent learning. The PBL activity accounted for 65 hours, which were divided into 26 class hours and 39 hours of personal and group work. Students carried out the project in small groups of three. The course was assessed via an exam weighing 25\% of their final grade, in addition to a design review (65\%) and a grade for attitude and participation (10\%). Again, student results were analysed using only nine questions, with no mention of the sample size, nor the number of students who took the survey.
\\
\vspace{5mm}
Similarly, team based learning (TBL) is another collaborative student centred teaching method, which was initially used for business teaching. Students are typically divided into groups of five to seven students to work on a project throughout the duration of a module. Unlike PBL, students are divided into teams of 5 to 7 people and are all facilitated by a single instructor \cite{Michaelsen2008, Parmelee2010, Michaelsen2011}. This is particularly advantageous in situations where faculty resources are limited and faculty members are required to supervise a growing number of students. Introductory pre-reading is required before each class in TBL, which is tested via two `readiness' tests. The first is an individual test, which is often a series of Multiple Choice Questions (MCQs), followed by a team test consisting of the same MCQs \cite{Burgess2018}. The motivation behind issuing the same MCQs to the teams is to promote team discussions. Subsequently, the last phase in the TBL activity involves issuing a team problem or task.
\\
\vspace{5mm}
However, each of these active learning techniques have their challenges in practically implementing them \cite{Guo2020}. In the case of TBL, developing multiple individual and team assessments before each class is time consuming. Furthermore, managing 320 students that are divided into 40-50 teams by a single instructor is simply infeasible. Moreover, for the effective implementation of PBL, developing a variety of engaging group projects in a single module is a technically demanding task for the instructors. Therefore, a combination of active learning techniques may be necessary, where multiple instructors are involved in coordinating teams of students who are given a single project to complete by the end of a course. This hybrid learning approach was suggested by Burgess \textit{et al.} in a medical curriculum, when a Problem Based Learning (PrBL) learning activity was converted to TBL and authors recommended a `hybrid [learning] approach utilising the strengths of both methods' \cite{Burgess2018}. In fact, Dolmans \textit{et al.} proposed a fusion between TBL and PrBL \cite{Dolmans2014} to optimise student learning in a medical curriculum. Similar to the medical profession, electronic engineering has witnessed a rapid transformation since the early invention of the transistor in 1947. Faced with this rapidly growing discipline, engineers must deal with uncertainty, incomplete data and a host of engineering problems with varying complexity. 
\\
\vspace{5mm}
While the differences between project and problem-based learning are indeed very subtle, a decision was made to combine the best aspects of project based learning with TBL. According to the literature, the main difference between problem and project-based learning is the deliverable that students submit to demonstrate their attainment of a set of learning outcomes \cite{Mills2003}. With problem-based learning, students are expected to develop a solution to a problem, whereas with project-based learning students are expected to deliver an entire product, service or process. Therefore, project-based learning is better suited to the environment that engineers typically work in. Furthermore, problem-based learning has its limitations in engineering education. Engineers must be able to use the knowledge gained from exposure to problems in their engineering education to be able to solve real world problems outside university \cite{mills2003engineering, Perrenet2000}. However, every problem will be different. Consequently, it may not be useful as an engineering practice to acquire knowledge that can be used and applied in practice. Moreover, since engineering education is hierarchical in nature, problem-based learning cannot be used to fill in missing gaps of knowledge. A detailed discussion regarding the differences between problem and project-based learning for engineering education is available in the literature \cite{Chen2020}.
\\
\vspace{5mm}
Consequently, since active learning has proven to effectively improve student learning, we designed a module that mixes both team-based and project-based learning, which can be effectively delivered to large student cohorts. Our motivation was to expose students to real-world authentic engineering problems, to increase overall student satisfaction and to develop their interpersonal skills as well as broad multidisciplinary knowledge. Our module was called Team Design and Project Skills (TDPS), which was delivered in the third year of an electronic engineering programme. From our previous investigations, we have compared student experiences from two different countries (UK and China) and found similarities in terms of appreciation for the work and satisfaction \cite{ghannam2020teachingteam}. Indeed, there were unpleasant feelings regarding student contributions to the group project, which led us to investigate the impact of Electronic Laboratory Notebooks (ELNs) on student learning and collaboration \cite{ghannam2020you, ghannam2021ELNs}. Our current work goes beyond existing literature in combining the best aspects of PBL and TBL in a single module that meets a variety of the UK`s Institution of Engineering and Technology (IET) graduate attributes, as evidenced from Table \ref{table:attributes}. We describe our approach in developing this module, how the tasks were aligned with accreditation body requirements, as well as student learning experiences during the past five years. 
\\
\vspace{5mm}
In the next section, the intended learning outcomes and the assessment methods are described. Next, the methodology for designing the open questionnaires to test student understanding and satisfaction are explained. In section four, student responses to the questionnaires are presented. Finally, concluding remarks and recommendations are provided in the final section of the paper. 

\section{Module Design}
As previously mentioned, real world engineering problems are often ill-bounded and ideal solutions do not exist. Moreover, the success of an engineering product is often determined by how well it achieves a budget and whether it was delivered on time \cite{Jonassen2006}. Thus, given a predefined budget, students in our module were encouraged to develop their own unique solutions. In fact, students were required to solve a variety of problems in this module. Among the 11 types of workplace engineering problems that have been identified by Johansen \cite{Jonassen2000}, the TDPS module aimed to expose students to the following types of problems: 

\begin{enumerate}
    \item Decision making – choosing between a limited range of alternatives 
    \item Troubleshooting – identifying possible faults in the hardware or software
    \item Planning – defining a management structure and plan for the team 
    \item Design – developing a rover that achieves these tasks.
\end{enumerate}

Furthermore, the TDPS module was worth 10 credits with an expected student workload of 100 hours at the undergraduate 3rd year level. In brief, the Intended Learning Outcomes (ILOs) of the module were: 

\begin{itemize}
    \item To adopt a `systems engineering' approach for the design and implementation of an electronic or technology-related product. 
    \item To take full responsibility for the complete project lifecycle, without relying on the instructor. 
    \item To gain and develop essential collaboration, management and leadership skills, as well as oral and written communication skills.
\end{itemize}

The learning activity involved developing a smart rover that can accomplish a set of communications, sensing and imaging tasks. Moreover, students were given a predefined budget (RMB1000) and a specific time-frame to complete the project (17 weeks). The module activities were designed to meet a diverse range of IET graduate attributes, as shown in Table \ref{table:attributes}.  Among the essential attributes that students must attain is the ability to ``work with information that may be incomplete or uncertain and be aware that this may affect the design''.
\\
\vspace{5mm}
Before the start of the module, a detailed module handbook was provided to the students and was uploaded on the university’s virtual learning environment (VLE), which was Moodle. The handbook contained detailed instructions regarding the task descriptions, the assessment mechanisms, the mark schemes and the assessment deadlines. In addition, the implementation guidelines (the rules) were provided. A mixture of interim and summative assessments were used to ensure student progress. There was no examination component in this module. This was to focus entirely on team based and project based learning. We also aimed to develop assessments that are similar and authentic to real-world engineering deliverables. The assessments were as follows:

\begin{itemize}
    \item Lab Notebook – This was used as an interim assessment to carefully monitor individual student progress. The notebook weighed 10\% of the final grade. 
    
    \item Oral Presentation – Instructors examined students to verify that the specifications have been met and to assess individual student contributions. This component weighed 25\% of the grade. 
    
    \item Live Demonstration – Practical live demonstration to evaluate and test the rover’s performance on a particular date. The live demonstration weighed 15\% of the final grade. 
    
    \item Team Report – Final documentation report, which weighed 25\% of the final grade. The report should explain the design, experimentation, testing and implementation of the technical product. 
    
    \item Individual Reflection Report – Report that highlights individual student contributions to the project. This report weighed 25\% of the final grade. 
    \end{itemize}

In the next section, the methods used to evaluate the effectiveness of the TDPS module are described.

\section{Methodology}
\subsection{Participants}
For the 2018/2019 academic year, 320 students enrolled in the TDPS module. The student cohort consisted of 84\% males and 26\% females. These students divided themselves into groups of eight, which meant that there were 40 groups of students. All these groups were moderated by three convenors. All teams were required to complete five assessments and the weighting of these assessments was previously mentioned in the module design section. Online surveys were administered to students in week 16 of the module to collect student feedback.

\subsection{Procedures}
\begin{figure}
  \includegraphics[width=\linewidth]{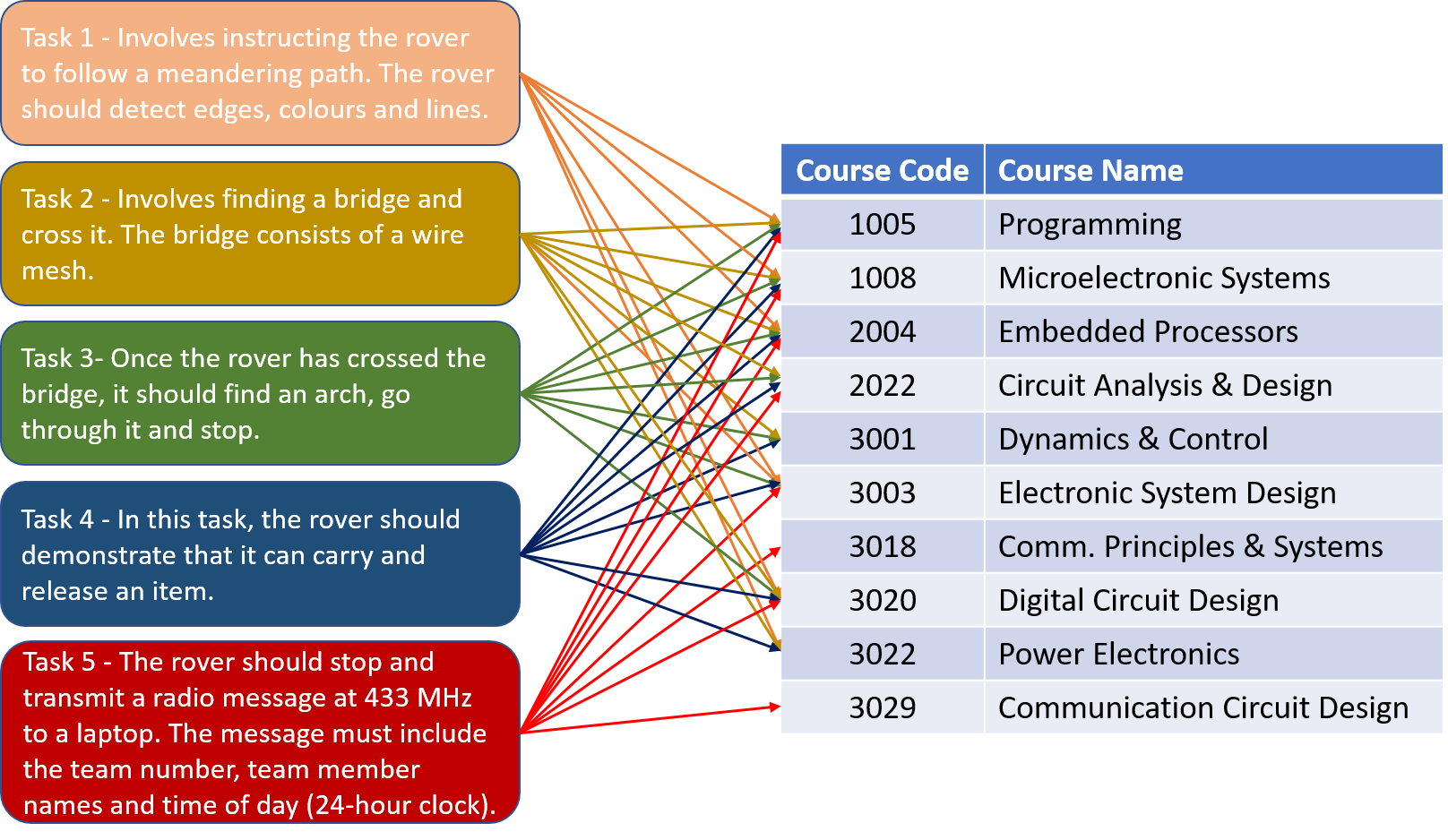}\\
  \caption{Technical tasks and their mapping to current modules in the XXXX programme. These modules are taught in the first, second and third years of the Electronic Engineering degree.}
  \label{fig:TaskMapping}
\end{figure}

\begin{figure}
  \includegraphics[width=\linewidth]{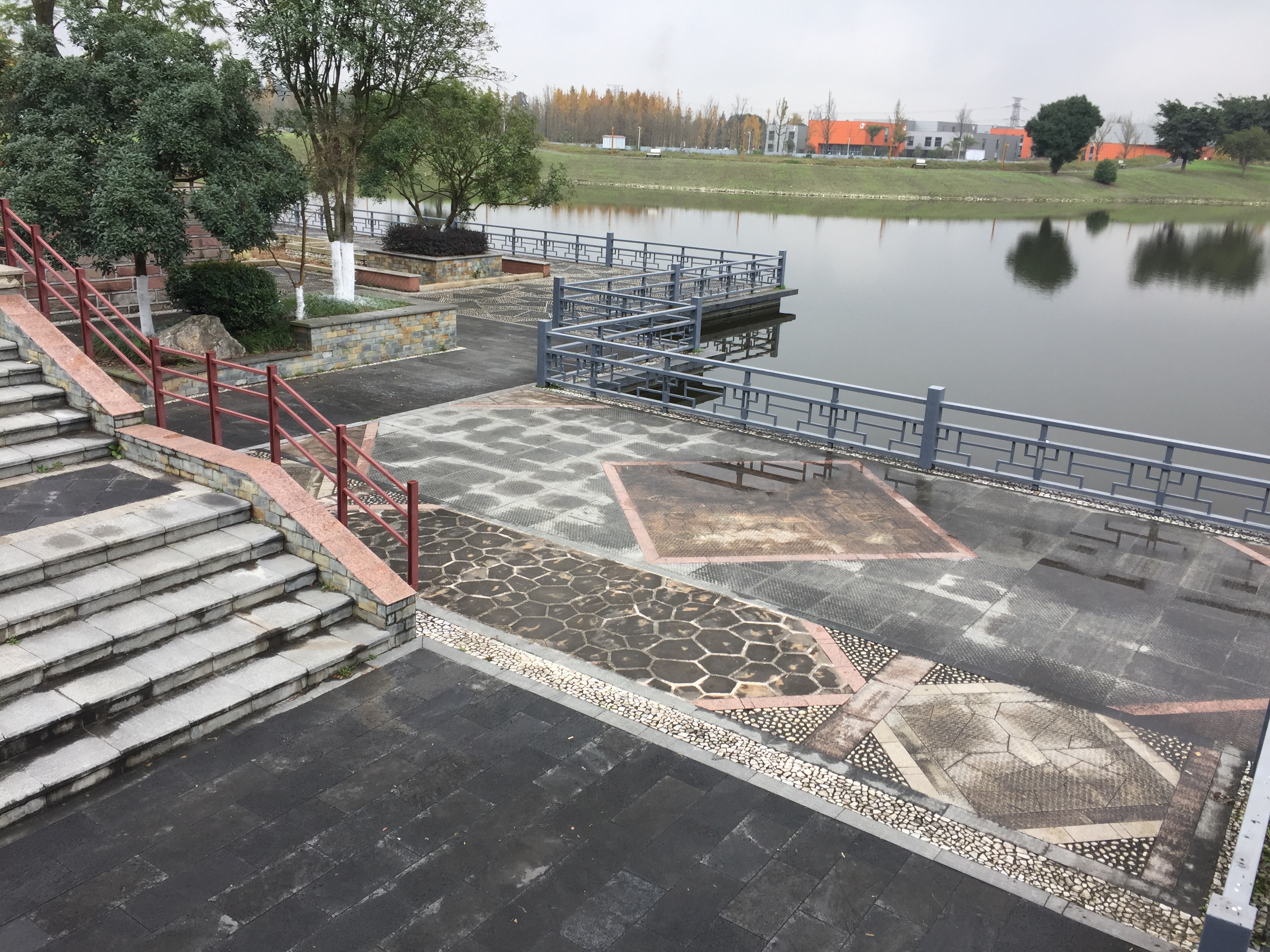}\\
  \caption{Track used for training the rovers at the University of Electronic Science and Technology of China (UESTC).}
  \label{fig:UESTC_Campus}
\end{figure}

The module was designed to enable students to acquire a set of technical and interpersonal skills as by products of the activities within the module. The technical tasks involved detecting colours, edges, lines, following a meandering path, carrying an item as well as transmitting a radio signal. These tasks were distributed within the patio shown in figure \ref{fig:UESTC_Campus}, which is located on the XXXX (XXXX) campus. A summary of the technical tasks and how these are mapped to other modules in the overall programme \cite{Ghannam2020Mapping} is shown in figure \ref{fig:TaskMapping}.
\\
\vspace{5mm}
Similar to the methodology adopted in \cite{Chen2017}, we evaluated the effectiveness of the module by gathering participant consent prior to undertaking this study. Students were informed that their participation was completely voluntary and that all collected information would be anonymous and confidential. They were also informed that they were able to withdraw their participation at any time. 

\subsection{Questionnaire Design}
Given the lack of research on this technology-enabled hybrid learning approach, we designed a questionnaire to gather student feedback regarding the effectiveness of this teaching approach. We therefore obtained the necessary ethical approvals from our College of Science and Engineering to distribute online surveys that consisted of 22 questions. These questions were divided into four sections, as shown in Table \ref{table:Questionnaire}. The first ten questions in Section 1 were designed to collect valuable student opinions regarding the overall quality of the module. Students were invited to indicate their learning experience via a 6-point Likert scale ranging from 0 (Disagree Entirely) to 5 (Agree Entirely) \cite{Carifio2007, Jamieson2004}. We preferred a 6-point scale to prevent students from giving “neutral” answers and to avoid student fatigue \cite{Leung2011}.
\\
\vspace{5mm}
Section 2 of the questionnaire consisted of three questions, which were concerned with obtaining feedback regarding the assessments used during the module. Section 3 consisted of five questions, which aimed to gather student feedback regarding their teamwork experience. Finally, section 4 of the questionnaire consisted of four questions, which were concerned with understanding how well the module met its ILOs.

\subsection{Data Analysis}
A total of 68 out of 320 students took part in our online survey, who provided useful feedback and recommendations. The response rate was 21.25\%, which easily surpassed the 8\% response rate deemed acceptable for a class size of 300 for a 10\% sampling error and 80\% confidence level \cite{Nulty2008}. A detailed analysis of the results obtained from these surveys is presented in the next section. 

\section{Results and Discussion}
By the end of the module, students were required to develop autonomously driven rovers that accomplished a certain set of tasks. Sample images of the rovers are shown in figure \ref{fig:RoverImages}. A sample video developed by the students is provided in the supplementary materials section.
\\
\vspace{5mm}
Results from the student responses to the questionnaires are shown in figures 5 to 9. According to the survey results presented in figure \ref{fig:Section1Data}, 59\% of students agreed that the module’s level of difficulty was appropriate for a third year module. None entirely disagreed. An area for further support and improvement could be in the number of contact hours of formal instruction. In their opinion, only six hours of lectures appeared insufficient, as shown in the results for Q3. However, our purpose was to move away from traditional lecture based instruction and towards independent learning. Perhaps this is attributed to the cultural background of our students. To remedy this problem, we propose more breakout or tutorial sessions for students to discuss their problems. We will also trial the use of technology, such as Piazza to enable greater interaction with students \cite{Ghannam2019a}.
\\
\vspace{5mm}
When asked whether this module helped improve their analytical and problem solving skills (Q4), 75\% of students either agreed, or strongly agreed. This clearly demonstrates that this module encouraged students to develop problem solving skills, enabling them to be better prepared for the real workplace.
\\
\vspace{5mm}
Moreover, students believed that this module broadened their interest in other fields of study (Q6). Almost 31\% of students strongly agreed. When asked for further details, students indicated image processing as the area most popular area. Other popular areas included project management and control systems, as shown from the data in figure \ref{fig:Otherfields}.
\\
\vspace{5mm}
Furthermore, 67\% of students believed that this module was a “valuable component in the degree programme”. In fact, 37\% strongly agreed that this was the case. Again, this reinforces previous findings in the literature that undergraduate students are generally in favour of active learning techniques \cite{Gallegos2011, Zhang2014, Zgheib2010, Alves2012}. Most importantly, 90\% of students believed that this module was useful for their future employment (Q9), with almost one third strongly agreeing that it will help them in the future. Similarly, 68\% of students either agreed or strongly agreed that they would recommend this module to future students (Q10). Only 3\% of students strongly disagreed, or disagreed that they would recommend this module to future cohorts. 
\\
\vspace{5mm}
There were areas for further improvement regarding the assessment mechanisms of the module (Q11). As evidenced from figure \ref{fig:Asssessments}, 82\% of students agreed that the range of assessment were appropriate for this module. Perhaps more can be done, such as the introduction peer assessment mechanisms. According to student responses to Q12 (c.f. figure \ref{fig:Asssessments}b), 69\% of students preferred to include peer assessment as part of the module’s assessment diet. Moreover, 58\% of students entirely or mostly agreed that that the range of assessments (oral presentation, demonstration, team report, lab report and individual report) were similar to real world or authentic engineering deliverables, as shown from the results in figure \ref{fig:Section2Data} (Q13).
\\
\vspace{5mm}
Another important aim of the module was the encouragement of teamwork. The module aimed to provide opportunities for collaborative learning, and to share knowledge in ways that are similar to the real world. Therefore, in section 3 of the questionnaire, students were asked a series of questions regarding teamwork and these results are shown in figure \ref{fig:Section2Data}. In fact, 86\% of students agreed that the module enabled them to work in a team to solve the given design problem (Q14). Furthermore, 31\% entirely agreed that working in a team helped students improve their problem solving skills (Q15). Most importantly, 43\% of student entirely agreed that working in a team helped “broaden their perspective in electronic system design” (Q16). 77\% of surveyed students agreed the module developed an understanding of team roles and for them to understand how to work effectively in a team (Q17).
\\
\vspace{5mm}
Finally, in section 4 of the questionnaire, it was important to understand how well the ILOs have been met. Students were asked whether they were now able to analyse technical requirements and to design, construct and test electronic hardware to perform specific functions. 95\% agreed that these ILOs have been met (Q18 and Q19). Other ILOs that perhaps needed further attention involve using project planning methodology to define milestones and measure achievement against such milestones (Q20). Students perhaps need further training in this area, especially from disciplines that involve operations research to become better familiar in project planning. Only a small minority (15\%) felt that this ILO has not been completely fulfilled. A similar minority (17\%) also felt that they were unable to “run a project without undue reliance on the instructor to perform productively as a team”. However, 70\% of students were able to “write a concise researched technical report that clearly addresses and analyses a particular issue or challenge”, as shown from the results in figure \ref{fig:Section2Data}.
\\
\vspace{5mm}
Next, student grades during the 5 years that the module was delivered are shown in figure \ref{fig:StudentGrades}. While there is skepticism regarding how well grades reflect student attainment, they are often used to reflect how well students have grasped a module's ILOs \cite{Schinske2014}. It is argued that higher grades indicate higher attainment and understanding of these ILOs. Accordingly, average grades for the TDPS module varied between 78 to 82\% during the past five years, as shown in figure \ref{fig:StudentGrades}. These average grades have been sustained, despite the increase in student enrollment numbers, which have almost trebled from 128 in year 1 to 320 in year 5. These student numbers are by far larger than those investigated by \cite{Martinez-Rodrigo2017, Calvo2018}. This clearly indicates that the module is popular among students. However, the standard deviation in student grades has been decreasing during the past 5 years. This shows that student grades are closely centered around the average grade, which also indicates that further work is required to ensure that assessments can distinguish between students with different abilities. Consequently, further work is required to improve the variability in student grades, which may be achieved by introducing peer-assessments.

\section{Conclusions}
Transitioning students from classical structured problems through worked examples to ill-defined problems is a subject of interest in the literature. This article demonstrates how a hybrid learning approach that combines Team Based and Project Based Learning has been used to design a new third year module called Team Design and Project Skills. The module was concerned with dividing 320 students into groups of eight in order to develop a rover that accomplishes a number of technical tasks. The motivation was to expose students to real-world authentic engineering problems and to develop some of the essential skills that are required by the continuously shifting job market. All students were given a predefined budget and were required to complete their products within 16 weeks. The module was exclusivity based on practical hands-on skills development and there was no exam component. Thus, the module has been well received during the past five years of instruction, with student numbers increasing from 128 to 320. A carefully designed survey consisting of 22 questions was used to probe student satisfaction. According to our surveys, which were completed by 68 students, 83\% admitted that the module was a valuable component in their degree programme. There were five assessment components, which involved presenting the project, demonstrating it, writing a technical report. Moreover, 82\% of the surveyed student agreed that these assessment were appropriate for the module. However, there are areas for further improvement. For example, students indicated a preference towards an increased number of lecture hours. To achieve this, supplementary online and interactive learning activities will be developed for students, which will be shared on the module's dedicated VLE.

\medskip
\medskip

\bibliographystyle{MSP}
\bibliography{MyLiterature}

\begin{thebibliography}{10}
\providecommand{\url}[1]{\texttt{#1}}
\providecommand{\urlprefix}{URL }

\bibitem{Kitchner1983}
K.~S. Kitchner,
\newblock \emph{Human Development} \textbf{1983}, \emph{26}, 4 222.

\bibitem{Jonassen2006}
D.~Jonassen, J.~Strobel, C.~B. Lee,
\newblock \emph{Journal of Engineering Education} \textbf{2006}, \emph{95}, 2
  139.

\bibitem{InternationalCo-operationEuropeLtd.ICEL2001}
B.~International Co-operation Europe~Ltd.(ICEL), Brussels,
\newblock \emph{Curriculum Development Guidelines. New ICT Curricula for the
  21st Century: Designing Tomorrow's Education},
\newblock ERIC Clearinghouse, \textbf{2001}.

\bibitem{Prince2004}
M.~Prince,
\newblock \emph{Journal of Engineering Education} \textbf{2004}, \emph{93}, 3
  223.

\bibitem{Freeman2014}
S.~Freeman, S.~L. Eddy, M.~McDonough, M.~K. Smith, N.~Okoroafor, H.~Jordt,
  M.~P. Wenderoth,
\newblock \emph{Proceedings of the National Academy of Sciences} \textbf{2014},
  \emph{111}, 23 8410.

\bibitem{Miller2014}
R.~Miller, J.~Euchner,
\newblock \emph{Research-Technology Management} \textbf{2014}, \emph{57}, 1 15.

\bibitem{baron2020adopting}
C.~Baron, B.~Daniel-Allegro,
\newblock \emph{Systems Engineering} \textbf{2020}, \emph{23}, 3 261.

\bibitem{Kokotsaki2016}
D.~Kokotsaki, V.~Menzies, A.~Wiggins,
\newblock \emph{Improving Schools} \textbf{2016}, \emph{19}, 3 267.

\bibitem{Macias-Guarasa2006}
J.~Macias-Guarasa, J.~Montero, R.~San-Segundo, A.~Araujo, O.~Nieto-Taladriz,
\newblock \emph{{IEEE} Transactions on Education} \textbf{2006}, \emph{49}, 3
  389.

\bibitem{Hosseinzadeh2012}
N.~Hosseinzadeh, M.~R. Hesamzadeh,
\newblock \emph{{IEEE} Transactions on Education} \textbf{2012}, \emph{55}, 4
  495.

\bibitem{Kellett2012}
C.~M. Kellett,
\newblock \emph{{IEEE} Transactions on Education} \textbf{2012}, \emph{55}, 3
  378.

\bibitem{Kumar2013}
A.~Kumar, S.~Fernando, R.~C. Panicker,
\newblock \emph{{IEEE} Transactions on Education} \textbf{2013}, \emph{56}, 4
  407.

\bibitem{Lee2010}
C.-S. Lee, J.-H. Su, K.-E. Lin, J.-H. Chang, G.-H. Lin,
\newblock \emph{{IEEE} Transactions on Education} \textbf{2010}, \emph{53}, 2
  173.

\bibitem{Lee2016}
M.~J.~W. Lee, S.~Nikolic, P.~J. Vial, C.~Ritz, W.~Li, T.~Goldfinch,
\newblock \emph{{IEEE} Transactions on Education} \textbf{2016}, \emph{59}, 4
  290.

\bibitem{Calvo2018}
I.~Calvo, I.~Cabanes, J.~Quesada, O.~Barambones,
\newblock \emph{{IEEE} Transactions on Education} \textbf{2018}, \emph{61}, 1
  21.

\bibitem{Hassan2015}
H.~Hassan, C.~Dominguez, J.-M. Martinez, A.~Perles, J.-V. Capella,
  J.~Albaladejo,
\newblock \emph{{IEEE} Transactions on Education} \textbf{2015}, \emph{58}, 3
  167.

\bibitem{Mantri2008}
A.~Mantri, S.~Dutt, J.~Gupta, M.~Chitkara,
\newblock \emph{{IEEE} Transactions on Education} \textbf{2008}, \emph{51}, 4
  432.

\bibitem{Salankar2021}
N.~Salankar, D.~Koundal, Y.-C. Hu,
\newblock \emph{Computer Applications in Engineering Education} \textbf{2021},
  1--15.

\bibitem{Durupinar2008}
F.~Durupinar, J.~Allbeck, N.~Pelechano, N.~Badler,
\newblock In \emph{Proceedings of the 7th International Joint Conference on
  Autonomous Agents and Multiagent Systems - Volume 3}, AAMAS '08.
  International Foundation for Autonomous Agents and Multiagent Systems,
  Richland, SC,
\newblock ISBN 9780981738123, \textbf{2008} 1217–1220.

\bibitem{Martinez-Rodrigo2017}
F.~Martinez-Rodrigo, L.~C. H.-D. Lucas, S.~de~Pablo, A.~B. Rey-Boue,
\newblock \emph{{IEEE} Transactions on Education} \textbf{2017}, \emph{60}, 3
  229.

\bibitem{Michaelsen2008}
L.~K. Michaelsen, M.~Sweet,
\newblock \emph{New Directions for Teaching and Learning} \textbf{2008},
  \emph{2008}, 116 7.

\bibitem{Parmelee2010}
D.~X. Parmelee, L.~K. Michaelsen,
\newblock \emph{Medical Teacher} \textbf{2010}, \emph{32}, 2 118.

\bibitem{Michaelsen2011}
L.~K. Michaelsen, M.~Sweet,
\newblock \emph{New Directions for Teaching and Learning} \textbf{2011},
  \emph{2011}, 128 41.

\bibitem{Burgess2018}
A.~Burgess, C.~Roberts, T.~Ayton, C.~Mellis,
\newblock \emph{{BMC} Medical Education} \textbf{2018}, \emph{18}, 1.

\bibitem{Guo2020}
P.~Guo, N.~Saab, L.~S. Post, W.~Admiraal,
\newblock \emph{International Journal of Educational Research} \textbf{2020},
  \emph{102} 101586.

\bibitem{Dolmans2014}
D.~Dolmans, L.~Michaelsen, J.~van Merriënboer, C.~van~der Vleuten,
\newblock \emph{Medical Teacher} \textbf{2014}, \emph{37}, 4 354.

\bibitem{Mills2003}
J.~E. Mills, D.~F. Treagust,
\newblock \emph{Australian Journal of Structural Engineering} \textbf{2003},
  \emph{4}, 3 211.

\bibitem{mills2003engineering}
J.~E. Mills, D.~F. Treagust, et~al.,
\newblock \emph{Australasian journal of engineering education} \textbf{2003},
  \emph{3}, 2 2.

\bibitem{Perrenet2000}
J.~C. Perrenet, P.~A.~J. Bouhuijs, J.~G. M.~M. Smits,
\newblock \emph{Teaching in Higher Education} \textbf{2000}, \emph{5}, 3 345.

\bibitem{Chen2020}
J.~Chen, A.~Kolmos, X.~Du,
\newblock \emph{European Journal of Engineering Education} \textbf{2020},
  \emph{46}, 1 90.

\bibitem{ghannam2020teachingteam}
R.~Ghannam, W.~Ahmad,
\newblock \emph{Compass: Journal of Learning and Teaching} \textbf{2020},
  \emph{13}, 2.

\bibitem{ghannam2020you}
R.~Ghannam,
\newblock \emph{IEEE Potentials} \textbf{2020}, \emph{39}, 5 21.

\bibitem{ghannam2021ELNs}
R.~Ghannam, S.~Hussain, H.~Fan, M.~{\'A}.~C. Gonz{\'a}lez,
\newblock \emph{IEEE Access} \textbf{2021}, \emph{9} 43241.

\bibitem{Jonassen2000}
D.~H. Jonassen,
\newblock \emph{Educational Technology Research and Development} \textbf{2000},
  \emph{48}, 4 63.

\bibitem{Ghannam2020Mapping}
R.~{Ghannam}, I.~S. {Ansari},
\newblock In \emph{2020 Transnational Engineering Education using Technology
  (TREET)}. \textbf{2020} 1--4.

\bibitem{Chen2017}
Z.~Chen,
\newblock \emph{{IEEE} Circuits and Systems Magazine} \textbf{2017}, \emph{17},
  1 33.

\bibitem{Carifio2007}
J.~Carifio, R.~J. Perla,
\newblock \emph{Journal of Social Sciences} \textbf{2007}, \emph{3}, 3 106.

\bibitem{Jamieson2004}
S.~Jamieson,
\newblock \emph{Medical Education} \textbf{2004}, \emph{38}, 12 1217.

\bibitem{Leung2011}
S.-O. Leung,
\newblock \emph{Journal of Social Service Research} \textbf{2011}, \emph{37}, 4
  412.

\bibitem{Nulty2008}
D.~D. Nulty,
\newblock \emph{Assessment {\&} Evaluation in Higher Education} \textbf{2008},
  \emph{33}, 3 301.

\bibitem{Ghannam2019a}
R.~Ghannam, H.~Qammer, S.~Hussain,
\newblock In \emph{Proceedings of the 7th Annual Conference of the UK \&
  Ireland Engineering Education Research Network}, Excellence in Engineering
  Education for the 21st Century: The Role of Engineering Education Research.
  WMG, University of Warwick, Coventry, UK, \textbf{2019} 33--41.

\bibitem{Gallegos2011}
P.~J. Gallegos, J.~M. Peeters,
\newblock \emph{Currents in Pharmacy Teaching and Learning} \textbf{2011},
  \emph{3}, 1 30.

\bibitem{Zhang2014}
D.~Zhang, N.~Yao, L.~Cuthbert, S.~Ketteridge,
\newblock In \emph{2014 IEEE Frontiers in Education Conference (FIE)
  Proceedings}. \textbf{2014} 1--8.

\bibitem{Zgheib2010}
N.~K. Zgheib, J.~A. Simaan, R.~Sabra,
\newblock \emph{Medical Teacher} \textbf{2010}, \emph{32}, 2 130.

\bibitem{Alves2012}
A.~C. Alves, D.~Mesquita, F.~Moreira, S.~Fernandes,
\newblock In \emph{Proceedings of the Fourth International Symposium on Project
  Approaches in Engineering Education (PAEE’2012),}, PAEE 2012. Paee
  Association, Braga, Portugal,
\newblock ISSN 9898525142, \textbf{2012} 23--32.

\bibitem{Schinske2014}
J.~Schinske, K.~Tanner,
\newblock \emph{{CBE}{\textemdash}Life Sciences Education} \textbf{2014},
  \emph{13}, 2 159.

\end{thebibliography}

\pagebreak
\clearpage

\begin{table}[h!]
	\small\centering
	\caption{Graduate attributes and their mapping to the TDPS module.}
    \begin{tabularx}{\textwidth}{ p{7.5cm}|p{11cm} }
       	\specialrule{2.5pt}{3pt}{3pt}
		IET Attributes & TDPS module Activity \\
		\specialrule{2.5pt}{3pt}{3pt}
		Understanding of, and the ability to apply, an integrated or systems approach to solving engineering problems. 
		&
		Students are expected to divide the system into component parts, with each team member taking responsibility for the analysis of each subsystem.\\
		\hline
		Plan and manage the design process, including cost drivers, and evaluate outcomes. 
		&
		Students are expected to manage the cost of the project and detailed design processes as a team, and are required to give a final oral presentation evaluating their final design. Students must manage the detailed design processes as a team, and are required to give a final oral presentation evaluating their design\\
		\hline
		Investigate and define the problem, identifying any constraints including environmental and sustainability limitations; ethical, health, safety, security and risk issues; intellectual property; codes of practice and standards. &
		Students working on the smart rover are expected to carry out their designs based on constraints discussed with other team members, within an overall budget, and (since the currents are moderately high) consider health and safety implications.\\
		\hline
		Knowledge and understanding of the scientific principles underpinning relevant current technologies, and their evolution. 
		&
		Practical knowledge of digital and analogue electronics, circuit theory and design. \\
		\hline
		Ability to apply quantitative methods in order to understand the performance of systems and components. 
		& 
		Students analyse the performance of their robots quantitatively and refine their design accordingly. \\
		\hline
		Ability to apply an integrated or systems approach to engineering problems through know‐how of the relevant technologies and their application. & 
		Practical implementation of basic electronics, programming, fault finding and testing. \\
		\hline
		Apply problem‐solving skills, technical knowledge and understanding to create or adapt design solutions that are fit for purpose. 
		& 
		Students are expected to design and fabricate a fully functional electronic system using their knowledge of microelectronics, embedded processors, power electronics, computer programming and other areas. Students must engineer their design solution without relying on their instructors. \\
		\hline
		Manage the design process, including cost drivers, and evaluate outcomes.
		& 
		Students are required to manage the design and implementation of a product under a fixed budget. \\ 
		\hline
		Work with information that may be incomplete or uncertain and be aware that this may affect the design.
		& 
		Uncertainty and variation are inherent in the design exercise. For example, there are intrinsic variations in the motors, batteries and electronics used by students to drive their autonomous vehicles. Students must therefore measure and judge the magnitude of these variations, and design with these uncertainties in mind. Students are not provided with detailed instructions for building their rovers. They must work in teams using incomplete information to design rovers that meet certain criteria.  \\
		\hline
		Demonstrate the ability to generate an innovative design for products, systems, components or processes to fulfil new needs. 
		&
		Practical experience in analogue and digital circuit design, and micro-controller code, requiring innovative solutions to novel problems, with the challenge changing each year.\\
		\hline
		Awareness of relevant legal requirements governing engineering activities, including personnel, health and safety,  intellectual property rights, product safety and liability issues
		&
		Students are expected to respect copyrights and other intellectual property. Students must also be aware of health and safety issues during fabrication of circuits with significant drive currents (fusing issues)\\
		\hline
		Communicate their work to technical and non‐technical audiences. 
		& 
		Students will demonstrate these via an oral presentation, a lab report and two written technical reports. \\ 
		\hline
		Knowledge of management techniques that may be used to achieve engineering objectives. 
		& 
		Project management tools such as Gantt and PERT charts as well as to explore the design cycle model. Students also have to complete the project within given budget, and sub-teams have to comply with internal deadlines to deliver the objectives of the project on time \\ 
		\hline
		Understanding of and ability to use relevant materials, equipment, tools, processes, or products. 
		& 
		Students must program software, use tools and various electronics products to assemble a smart rover \\ 
		\hline
		Ability to use and apply information from technical literature.
		& 
		Students must carefully examine data sheets and technical literature before purchasing and integrating equipment, which is required for designing and implementing a robotic product. \\ 
		\hline
		Ability to use appropriate codes of practice and industry standards.
		& 
		Students learn to use equipment that meets industry standards. \\ 
		\hline
		Understanding of the use of technical literature and other information sources
		&
		Extensive use of technical data sheets, including parsing of key information to make component choice decisions \\
		\hline
		Awareness of team roles and the ability to work as a member of an engineering team.
		& 
		Students must work in a team to achieve the desired results. They learn how to divide activities and achieve deliverables within a specific time-frame\\ 
		\specialrule{2.5pt}{3pt}{3pt}
	\end{tabularx}
	\label{table:attributes}
\end{table}

\begin{table}[h!]
    \small\centering
	\caption{Survey questions used to evaluate the effectiveness of our hybrid teaching approach that aimed to develop `systems engineers'.}
    \begin{tabularx}{\textwidth}{ p{1.5cm}|p{17cm} }
       	\specialrule{2.5pt}{3pt}{3pt}
		Question & Description \\
        \specialrule{2.5pt}{3pt}{3pt}
		Q1 &
		Do you feel that the level of difficulty was appropriate for third-year undergraduate study? \\
		Q2 &
		Do you feel that this module helped you understand how to deal with complex engineering problems? \\
		Q3 & 
		Was enough lecture material to guide the learning process? \\
		Q4 & 
		Do you believe that this module helped improve your analytical and problem solving skills? \\
		Q5 & 
		Do you feel that this module helped broaden your perspective in the area of electronic system design? \\ 
		Q6 & 
		Do you believe that this module broadened your interest in other areas of study? If so, please indicate what areas? \\ 
		Q7 & 
		Do you feel that this module was a valuable component in the degree programme? \\ 
		Q8 & 
		Do you feel that the learning experience from this module will benefit your final year project? \\
		Q9 & 
		Do you believe that this module will be useful for your future employment? \\ 
		Q10 &
		Would you recommend this module to your colleagues? 
 \\
		\hline
		 Q11 & 
		 Do you believe that the range of assessment mechanisms were appropriate for this module? \\
		 Q12 & 
		 Do you believe that the peer assessment should also be included as part of the assessment mechanism in this module? (Peer assessment involves students taking responsibility for assessing the work and performance of their peers against set assessment criteria.) \\ 
		 Q13 & 
		 Do you believe that the assessments were similar to real world or authentic engineering deliverables? \\
		\hline
		Q14 & 
		Do you believe that working in a team helped you solve the design project given in this module? \\
		Q15 & 
		Do you believe that working in a team helped improve your problem-solving skills? \\
		Q16 & 
		Do you believe that working in a team helped broaden your perspective in electronic system design? \\
		Q17 & 
		Has the module developed your awareness of team roles and your ability to work as a member of an engineering team? \\
		Q18 & 
		Analyse technical requirements to develop an overall design plan. \\
		\hline
		Q19 & 
		Design, construct and test electronic hardware to perform specific functions. \\ 
		Q20 & 
		Use a project planning methodology (such as Gantt charts) to define milestones and measure achievement against such milestones. \\ 
		Q21 & 
		Run a project without undue reliance on the module instructor to perform productively as a team and to recognise contributions from all team members. \\
		Q22 & 
		Write a concise researched technical report that clearly addresses and analyses a particular issue or challenge. \\
		\hline
	\end{tabularx}
	\label{table:Questionnaire}
\end{table}

\clearpage
\newpage

\begin{figure}
  \includegraphics[width=\linewidth]{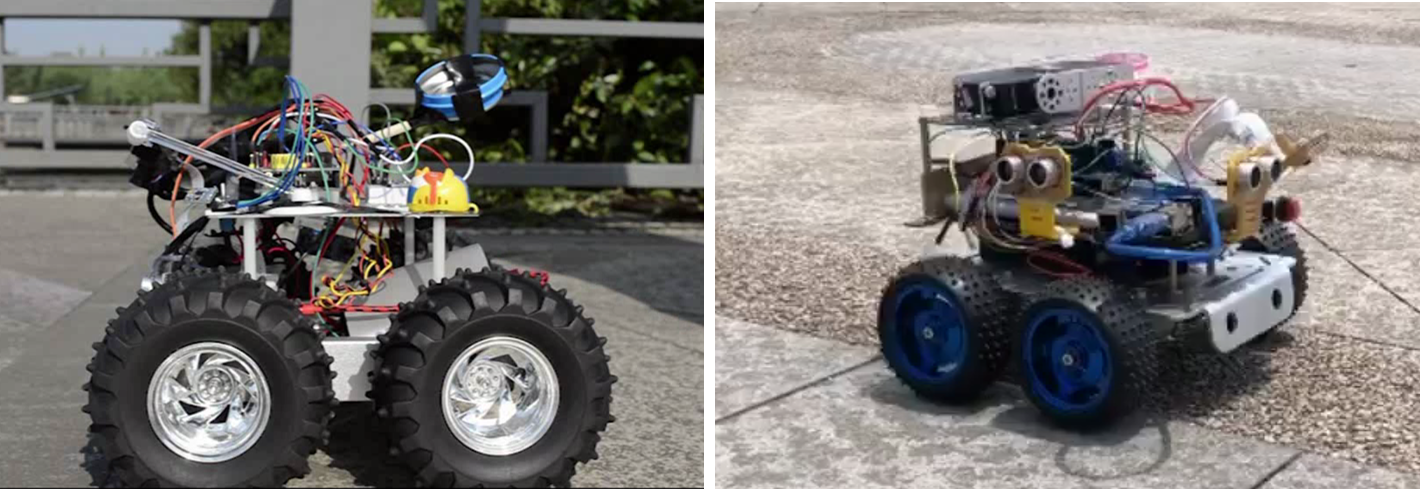}\\
  \caption{Sample images of the rovers developed by the teams of students. These rovers were designed to detect images, lines, transmit information and carry an item.}
  \label{fig:RoverImages}
\end{figure} 

\begin{figure*}
  \includegraphics[width=\linewidth]{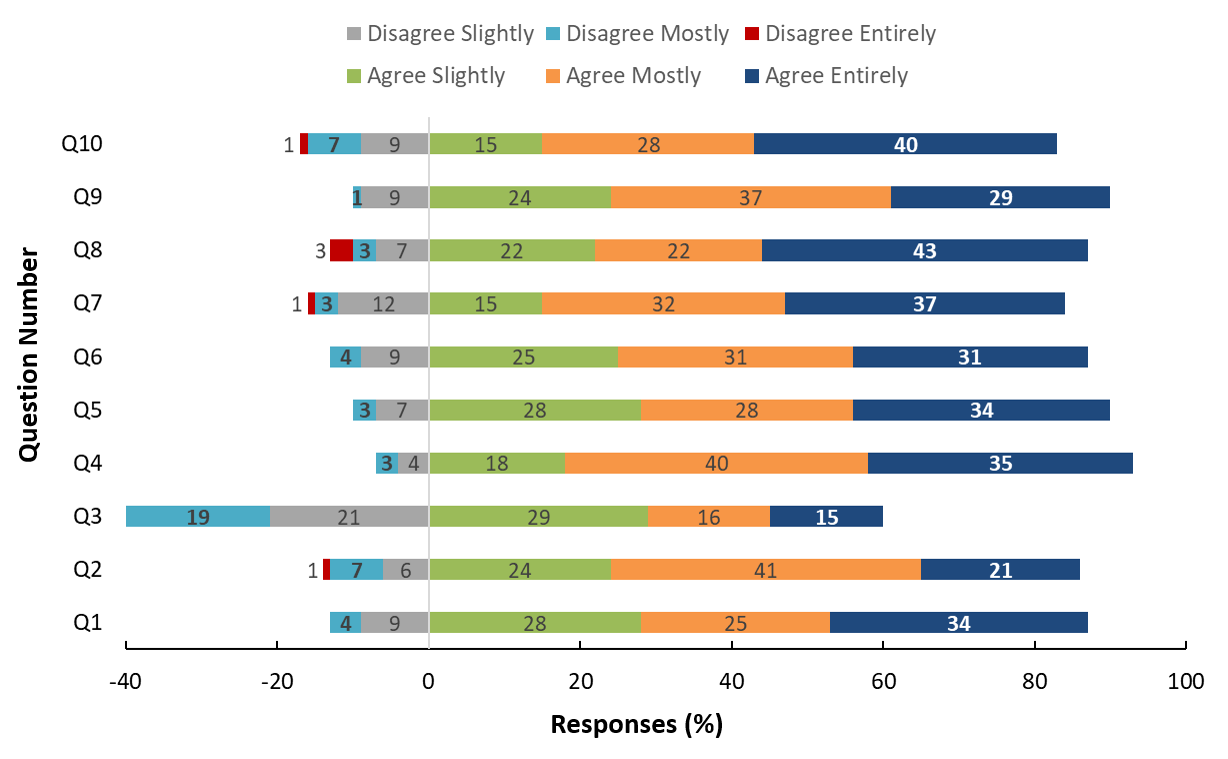}\\
  \caption{Student responses to Section 1 survey questions. The first 10 questions were: Q1: Do you feel that the level of difficulty was appropriate for third-year undergraduate study? Q2: Do you feel that this module helped you understand how to deal with complex engineering problems? Q3: Was enough lecture material to guide the learning process? Q4: Do you believe that this module helped improve your analytical and problem solving skills? Q5: Do you feel that this module helped broaden your perspective in the area of electronic system design? Q6: Do you feel that this module broadened your interest in other areas of study? If so, please indicate which areas? Q7: Do you feel that this module was a valuable component in the degree programme? Q8: Do you feel that the learning experience from this module will benefit your final year project? Q9: Do you believe that this module will be useful for your future employment? Q10: Would you recommend this module to your colleagues?}
  \label{fig:Section1Data}
\end{figure*} 

\begin{figure}
  \includegraphics[width=\linewidth]{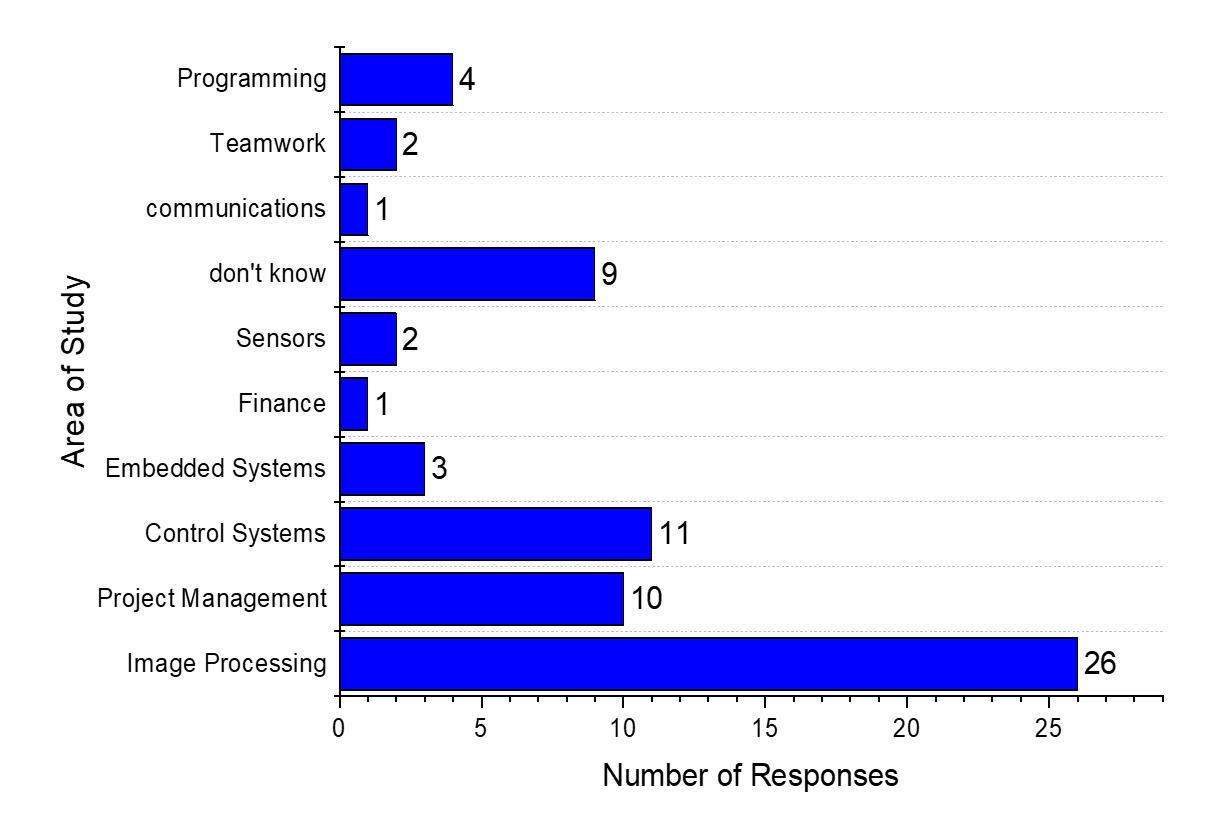}\\
  \caption{Areas of further study that have been learned during this module. Vast majority of students mentioned that Image Processing, Project Management and Control Systems were the most popular areas of further study.}
  \label{fig:Otherfields}
\end{figure} 

\begin{figure}
  \includegraphics[width=\linewidth]{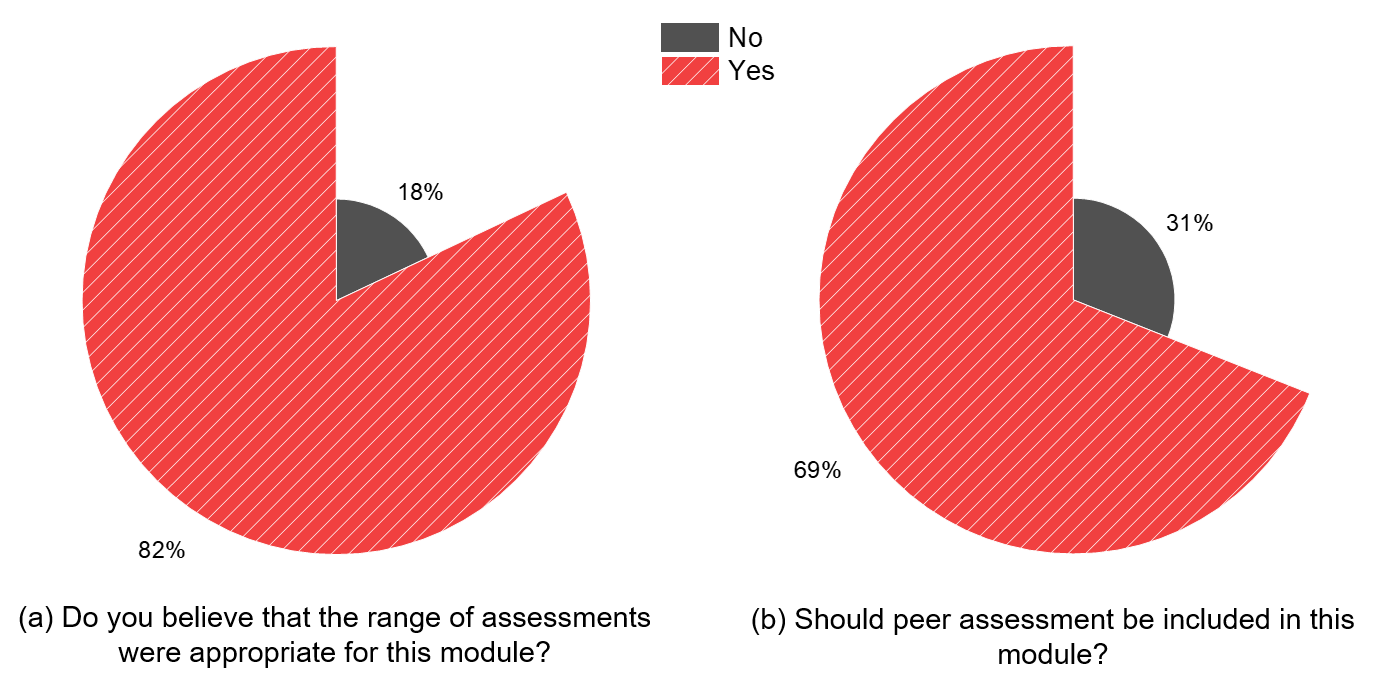}\\
  \caption{Students responses to (a) question 11 and (b) question 12 in the survey. The majority of students were clearly satisfied with the range of assessments, but felt that peer assessment should be included in the future.}
  \label{fig:Asssessments}
\end{figure} 

\begin{figure*}
  \includegraphics[width=\linewidth]{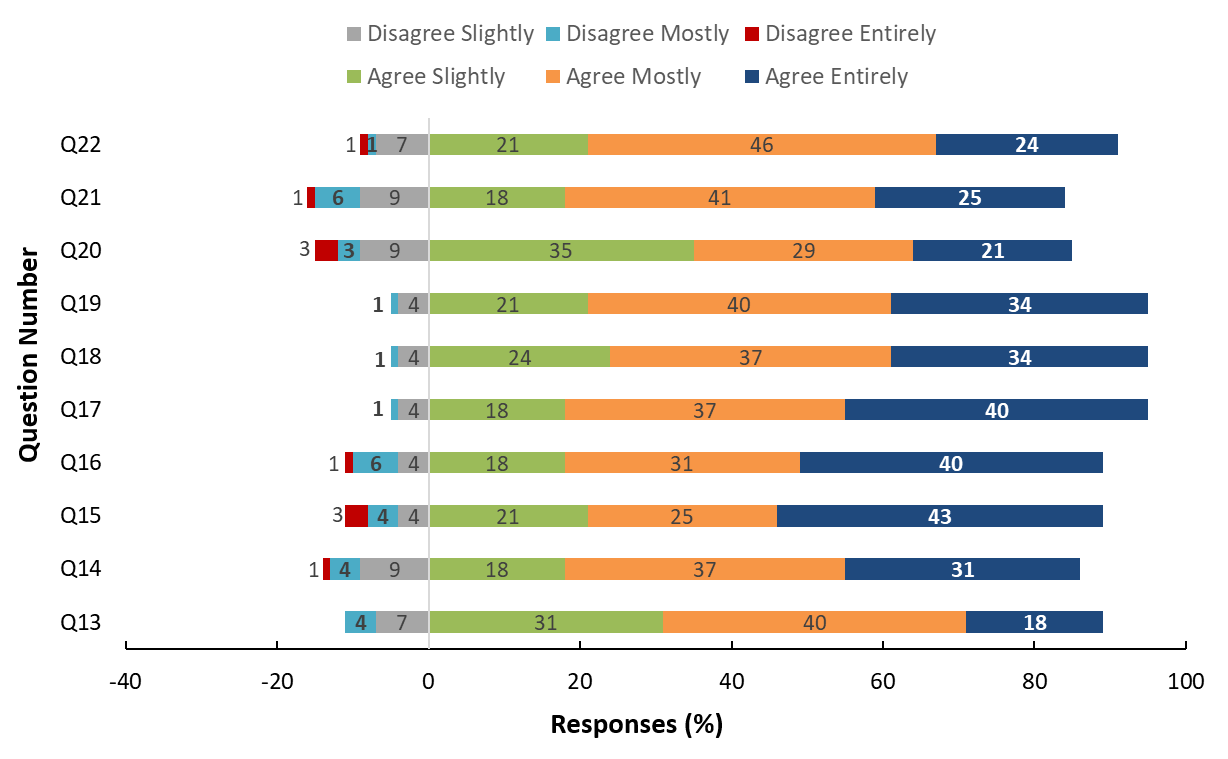}\\
  \caption{Student responses the survey questions. The questions were: Q13: Do you believe that the assessments were similar to real world or authentic engineering deliverables? Q14: Do you believe that working in a team helped you solve the design problem given in this module? Q15: Do you believe that working in a team helped improve your problem-solving skills? Q16: Do you believe that working in a team helped broaden your perspective in electronic system design? Q17: Has the module developed your awareness of team roles and your ability to work as a member of an engineering team? Q18: Has the module enabled you to analyse technical requirements to develop an overall design plan. Q19: Has the module enabled you to design, construct and test electronic hardware to perform specific functions. Q20: Has the module enabled you to use a project planning methodology (such as Gantt charts) to define milestones and measure achievement against such milestones. Q21: Run a project without undue reliance on the instructor to perform productively as a team. Q22: Write a concise researched technical report that clearly addresses and analyses a particular issue or challenge. }
  \label{fig:Section2Data}
\end{figure*} 

\begin{figure}
  \includegraphics[width=\linewidth]{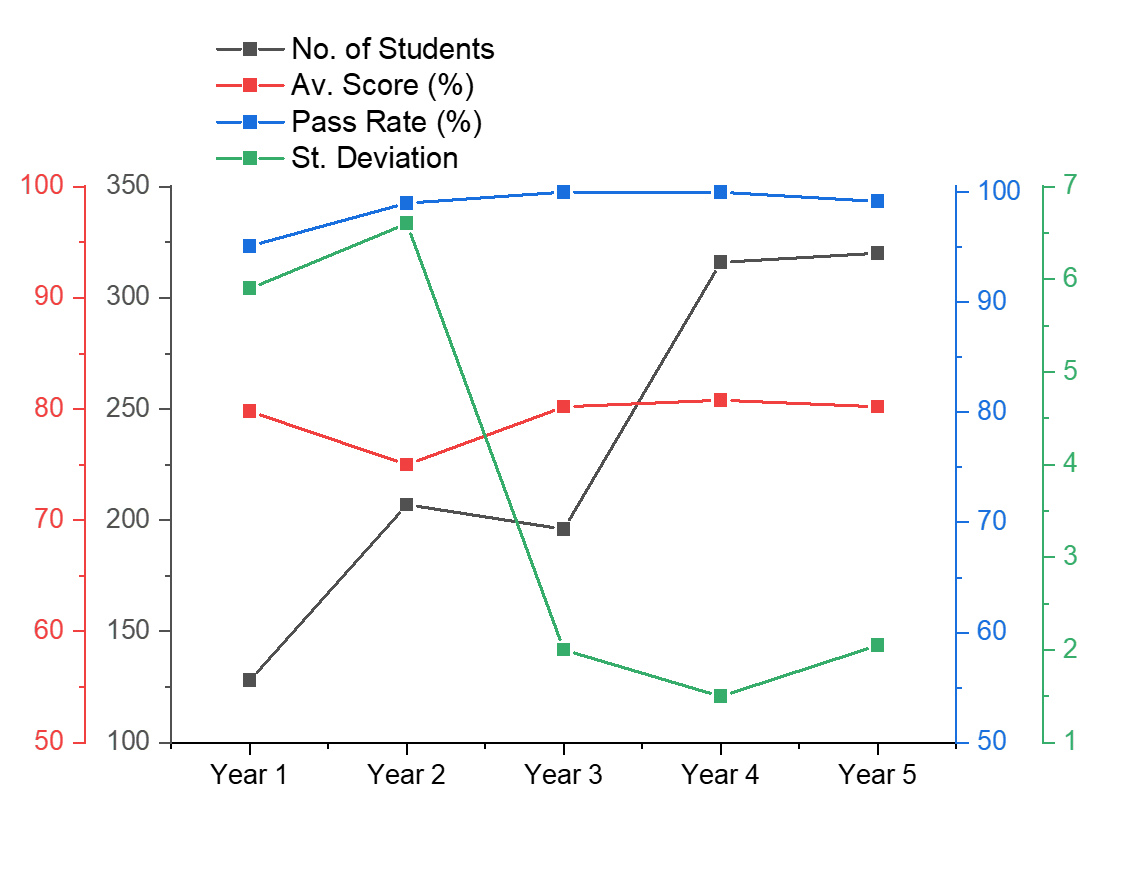}\\
  \caption{Student statistics for the TDPS module during the past five years. The graph shows average number of students, average scores, the pass-rate and the standard deviation in student grades. All statistics are positive, the the exception of the standard deviation. Further work is therefore required to distinguish between the excellent and the weaker students.}
  \label{fig:StudentGrades}
\end{figure}

\clearpage
\pagebreak[4]
\newpage

\end{document}